# Spin-Hall Nano-oscillator: a micromagnetic study

A. Giordano,[1] M. Carpentieri,[2] A. Laudani,[3] G. Gubbiotti,[4] B. Azzerboni,[1] G. Finocchio[1]

[1] Department of Electronic Engineering, Industrial Chemistry and Engineering. University of Messina, C.da di Dio, I-98166, Messina, Italy.

[2] Department of Electrical and Information Engineering, Politecnico of Bari, via E. Orabona 4, I-70125 Bari, Italy.

[3] Department of Engineering, University of Roma Tre, via V. Volterra 62, I-00146 Roma, Italy.

[4] Istituto Officina dei Materiali del CNR (CNR-IOM), Unità di Perugia c/o Dipartimento di Fisica, Via A. Pascoli, 06123, Perugia, Italy

**Abstract:** This letter studies the dynamical behavior of spin-Hall nanoscillators from a micromagnetic point of view. The model parameters have been identified by reproducing recent experimental data quantitatively. Our results indicate that a strongly localized mode is observed for in-plane bias fields such as in the experiments, while predict the excitation of an asymmetric propagating mode for large enough out-of plane bias field similarly to what observed in spin-torque nanocontact oscillators. Our findings show that spin-Hall nanoscillators can find application as spin-wave emitters for magnonic applications where spin waves are used for transmission and processing information on nanoscale.

Experimental studies of bilayer composed by a heavy metal film coupled with a thin ferromagnet have opened a route on the development of a more efficient category of spintronic devices where the magnetic state can be changed by the effects related to spin-orbit coupling, such as Dzyaloshinskii-Moriya interaction, Rashba and spin-Hall effects.[1,2,3,4,5,6,7] In particular, domain wall motion at very high velocity,[3] magnetization reversal[1,2,8,9,] and persistent magnetization precession have been achieved.[10,11,12] These results are motivated by technological interest, aimed to design the next generation of nanomagnetic logic gates, magnetic memories and nanoscale oscillators. In this letter, we focus on this latter device category. The persistent magnetization precession driven by spin-orbit interactions in Pt/Py bilayer was first measured by Demidov *et al*[10] where the torque from spin-orbit coupling, mainly spin-Hall effect, originating from a bias current flowing in the Pt layer was large enough to compensate the magnetic losses and to excite a persistent magnetization precession in the Py layer. Demidov *et al*[10] also demonstrated the nature of the excited mode to be a non-propagating spin wave with localization region of less than 100 nm. Here we performed a systematic study of that experimental system[10] to understand the physical origin of the excited modes. For in plane-fields, our computations reproduce quantitatively the experimental oscillation frequency as a function of current and the localization of the excited mode. For out-of-plane-fields, we predict the excitation of propagating spin waves with an asymmetric propagation pattern. Our findings show this device geometry is prototype for spin-wave emitters for magnonic applications where spin waves are used for transmission and processing information on nanoscale.[13,14]

The device is a bilayer composed by Pt(8)/Py(5) (thicknesses in nm). The bias current is injected in the center of the Py layer by means of two triangular Gold (Au) contacts (thickness of 150 nm) at a nominal distance *d*. A sketch of the device is displayed in Fig. 1(a). A Cartesian coordinate system is introduced, where the *x*-axis is parallel to the direction of the injected current, while the *y* and *z* axes are the other in-plane and the out-of-plane directions, respectively. The in-plane field has been applied along the *y*-direction to saturate the magnetization along that direction. The out-of-plane field has been applied in the *yz*-plane with a 15° angle with respect to the z-axis. In general, the magnetization dynamics should be studied by the Landau-Lifshitz-Gilbert equation which takes into account: the adiabatic $\tau_{A-ST}$ and non-adiabatic $\tau_{NA-ST}$ spin-transfer-torques due to the current which flows into the ferromagnet, the spin-orbit torques from the spin Hall $\tau_{SHE}$ and the Rashba $\tau_{RE}$ effects:[8, 15,16,17,18,19]

$$\frac{d\mathbf{M}}{dt} = -\gamma_0 \mathbf{M} \times \mathbf{H}_{\mathbf{EFF}} + \frac{\alpha}{M_S} \mathbf{M} \times \frac{d\mathbf{M}}{dt} + \tau_{SHE} + \tau_{RE} + \tau_{A-ST} + \tau_{NA-ST} \qquad (1)$$

where $\mathbf{M}$ and $\mathbf{H}_{\mathbf{EFF}}$ are the magnetization and the effective field vectors of the ferromagnet, respectively. $\mathbf{H}_{\mathbf{EFF}}$ takes into account the exchange, the magnetocrystalline anisotropy, the external, the self-magnetostatic fields and the Oersted field. $\alpha$, $M_S$ and $\gamma_0$ are the Gilbert damping, the saturation magnetization, and the gyromagnetic ratio, respectively. The first step is the computation of the spatial distribution of the current density in the Py and Pt layers by means of the Ohm's law $\mathbf{J} = \rho^{-1}\mathbf{E}$ where $\mathbf{E}$ is the local electric field and ρ is the resistivity of the material. The electric field can be computed as the gradient field of the electrostatic potential *V*, so that $\mathbf{E} = -\nabla V$. For the charge conservation law that yields $\nabla \cdot \mathbf{J} = 0$, we have $\nabla \cdot (\rho^{-1} \nabla V) = 0$. This kind of Elliptic differential equation is numerically solved with the Finite Element Method (1st order tetrahedral, employing more than $3 \times 10^6$ elements and about $6.7 \times 10^5$ nodes)[20] with the boundary condition $\partial V / \partial n = 0$ (where *n* is the normal to the geometrical boundary) except for the contacts, where the current density is injected.

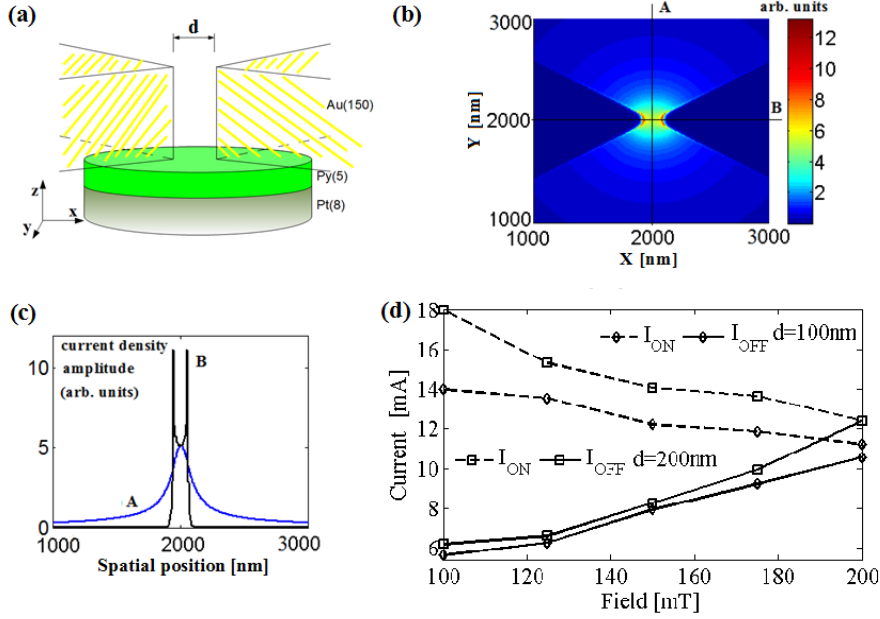

FIG. 1 (color online): (a) Schematic view of the studied device, $d$ is the distance between the Au contacts. Thicknesses of the layer are expressed in nm. The coordinate axes are also shown. (b) Example of spatial distribution of the current density as computed numerically (reduced square region of 2000x2000nm$^2$ of a disk of 4μm in diameter), (c) current profiles for the sections A and B as indicated in (b). (d) Critical currents $I_{OFF}$ and $I_{ON}$ as a function of the applied field for two different electrodes distances for $d$=100 (solid and dotted lines with circles) and 200nm (solid and dotted lines with squares).

Figure 1(b) shows a reduced region (between 1000nm and 3000nm) of the spatial distribution of the current density in the Pt layer as computed for a disk of 4 μm in diameter, the material conductivities used for the computation are 4.1x10$^8$ (Ωm)$^{-1}$, 5.1x10$^7$ (Ωm)$^{-1}$ and 6.4x10$^6$ (Ωm)$^{-1}$ for the Gold, Platinum, and Permalloy, respectively. The current flows almost totally in the two Gold contacts up to the position where it is injected in the Py/Pt bilayer. In between the two contacts, around 90% of the current flows in the Pt layer. In terms of spatial distribution, the current is localized mainly in the center of the system with a symmetric spread around the $y$-direction (perpendicular to an ideal connection line of the two contacts). The current profile for two particular sections (A blue line and B black line) of the device are displayed in Fig. 1(c). In the Section A (*y*-direction perpendicular to the current flow) the current density exhibits a maximum value in the center. On the other hand, the results related to the Section B (parallel to the current flow) indicate the presence of two maxima near the boundary with the two contacts at the position where the current starts and stops the flow in the Py/Pt bilayer.

The $\tau_{A-ST}$, $\tau_{NA-ST}$ and $\tau_{RE}$ torques, being proportional to the current density flowing into the ferromagnet (less than 10%), are negligible with respect to the $\tau_{SHE}$, which is proportional to the current which flows into the Pt layer (more than 90%).[21] With this in mind, Eq. (1) can be simplified as follows:

$$\frac{d\mathbf{M}}{dt} = -\gamma_0 \mathbf{M} \times \mathbf{H}_{\mathbf{EFF}} + \frac{\alpha}{M_S} \mathbf{M} \times \frac{d\mathbf{M}}{dt} - \frac{\mu_B \alpha_H J_{Pt}(x,y)}{eM_S^2 t_{Py}} \mathbf{M} \times \mathbf{M} \times \mathbf{\sigma} \qquad (2)$$

where $J_{Pt}(x,y)$ is the spatial distribution of the modulus of the current density in the Pt layer considering the same sign of the applied current, $\mu_B$ the Bohr Magneton, $e$ the electric charge, and $t_{Py}$ the thickness of the Py-layer. $\mathbf{\sigma}$ is the polarization of the spin current due to the spin-dependent scattering in the Pt layer. For each computational cell, the current density vector $\mathbf{J}_{Pt}$, $\mathbf{\sigma}$, and the *z*-axis satisfy the relationship $\mathbf{\sigma} = \mathbf{z} \times \frac{\mathbf{J}_{Pt}}{|\mathbf{J}_{Pt}|}$.[4] $\alpha_H$ is the spin Hall angle given by the ratio between the amplitude of the transverse spin current density generated in the Pt and the charge current density flowing in it. The parameters used for our numerical simulations are: exchange constant $1.3 \times 10^{-11}$ J/m, spin-Hall angle 0.08, Gilbert damping 0.02, and saturation magnetization $650 \times 10^3$ A/m.[25] While the current density distribution and the Oersted field have been computed by considering a disk with a diameter of 4μm, the micromagnetic computations have been performed for a disk diameter of 2.5μm to reduce the computational time of the systematic study.[22] The cubic discretization cell is 5nm in side, which is smaller than the exchange length for the Py ($\approx$ 7nm). The effects of the thermal fluctuations have been taken into account as a random thermal field $\mathbf{H}_{th}$ added to the effective field for each computational cell.[23, 24, 25]

The first step of our analysis is to understand the origin of the persistent magnetization precession measured experimentally by Demidov *et al.*[10] A systematic study has been performed as a function of the current and the magnetic field applied along the *y*-direction and for different distances *d* between the two Au contacts (see Fig. 1a). For each magnetic field value, the initial configuration of the magnetization has been computed by solving the Eq. 2 with $J_{Pt}(x,y)$ equal to zero up to reach the condition that $\mathbf{M}$ is parallel to $\mathbf{H}_{\mathbf{EFF}}$ ($\frac{d\mathbf{M}}{dt} \approx 0$). Starting from that static state, the dynamical response of the magnetic device has been computed by sweeping the current back and forth from 0 up to 20mA. For increasing current, a critical value $I_{ON}$ exists where the magnetic configuration becomes unstable and a self-oscillation is then excited. On the other hand, for decreasing current the self-oscillation is switched off at $I_{OFF} < I_{ON}$. Fig. 1(d) summarizes $I_{OFF}$ and $I_{ON}$ as a function of the external field for *d*=100nm (solid and dotted lines with square) and 200 nm (solid and dotted line with circles). At *d*=200nm and for the whole range of the applied field, both $I_{OFF}$ and $I_{ON}$ are larger than the one at *d*=100nm. This is because at *d*=200nm the spatial distribution of the current density is widely spread when compared to the configuration at *d*=100nm and consequently it should be larger to compensate the losses due to the Gilbert damping. We also computed $I_{ON}$ for *H*=100mT at *d*=150nm finding a value of 16mA which is in between the ones at *d*=100nm (14mT) and at *d*=200nm (18mT).

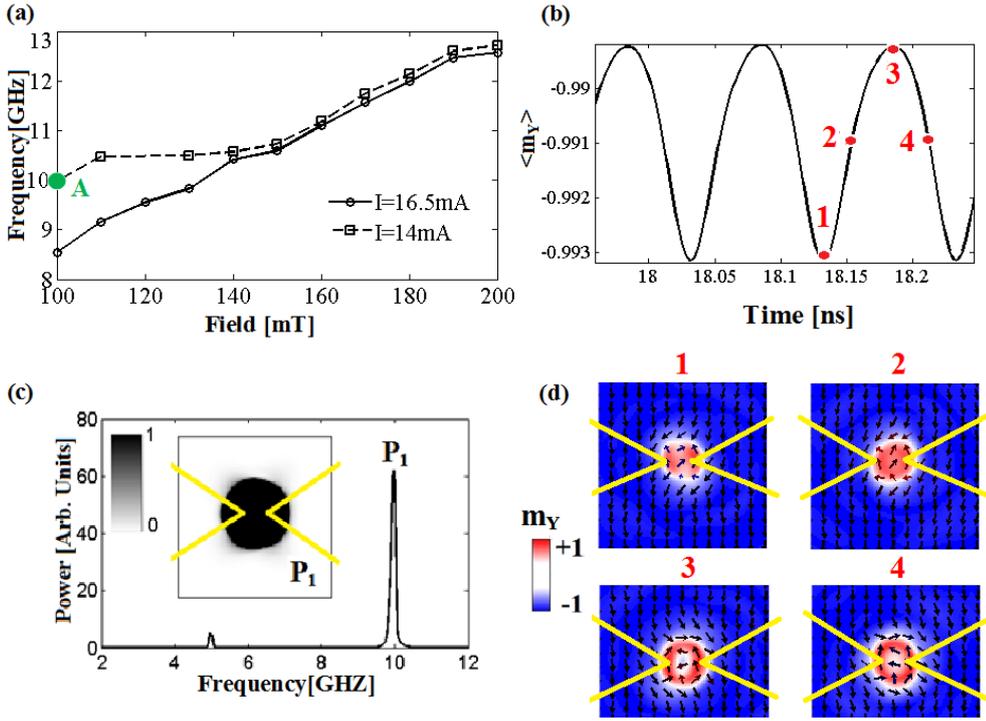

FIG.2: (color online) (a) Oscillation frequencies as a function of the applied field for $I$=14mA (dotted line with squares) and $I$=16.5mA (solid line with circles) at $d$=100 nm. (b) Time domain trace of the average $y$-component of the normalized magnetization computed for the whole cross section of the Py layer for $I$=14mA and $H$=100mT. (c). Main panel: Power spectrum of the self oscillation excited at $I$=14mA, $H$=100mT as computed with the micromagnetic spectral mapping technique. Inset: spatial distribution of the excited mode $P_1$ (the power increases from white to black, the yellow lines indicate the location of the Gold contacts). (d) Snapshots of the magnetization (reduced square region of $500\times500nm^2$) as computed at the time indicated with the points 1-4 in Fig. 2(b) (the color is related to the $y$ component of the magnetization $m_Y$, while the arrows indicate the in-plane component of the magnetization). (multimedia view).

Fig. 2(a) summarizes the oscillation frequencies ($d$=100nm) for two currents ($I$=14 and 16.5 mA) as a function of the field. At fixed field the oscillation frequency exhibits red shift similarly to what observed for STT oscillators and as expected for self-oscillations with an in-plane oscillation axis.[26] The dynamical state computed at $H$=100mT, $d$=100nm and $I$=14mA is described in detail (point A of Fig. 2(a)), but qualitative similar results are observed for other currents. Fig. 2(b) shows the temporal evolution of the *y*-component of the spatial average of the normalized magnetization precession as computed for the whole Py-layer. As can be observed, the magnetization oscillation is only 2% of the possible maximum oscillation and it is localized near the gold contact region, see a zoom of the snapshots in Fig.2(d) related to the points 1-4 as indicated in Fig. 2(b). At the critical current $I_{ON}$, the magnetization close to the gold contacts starts to oscillate around an oscillation axis which is reversed (along -*y*-direction) with respect to the equilibrium configuration (+*y*-direction). Fig. 2(c) (main panel) shows the power spectrum of the self-oscillation achieved for $H$=100mT, $d$=100nm and $I$=14mA as computed with the micromagnetic spectral mapping technique.[27, 28] The excited mode **P₁** is characterized by an oscillation frequency of 9.98GHz and a uniform spatial distribution, as displayed in the inset of Fig. 2(c). The mode is strongly localized in the central region of the device where the current is injected into the bilayer. This result reproduces the experimental finding of the spatial distribution of the mode measured in[10] (compare the inset of Fig. 2(c) with Fig. 3 of Ref. [10]).

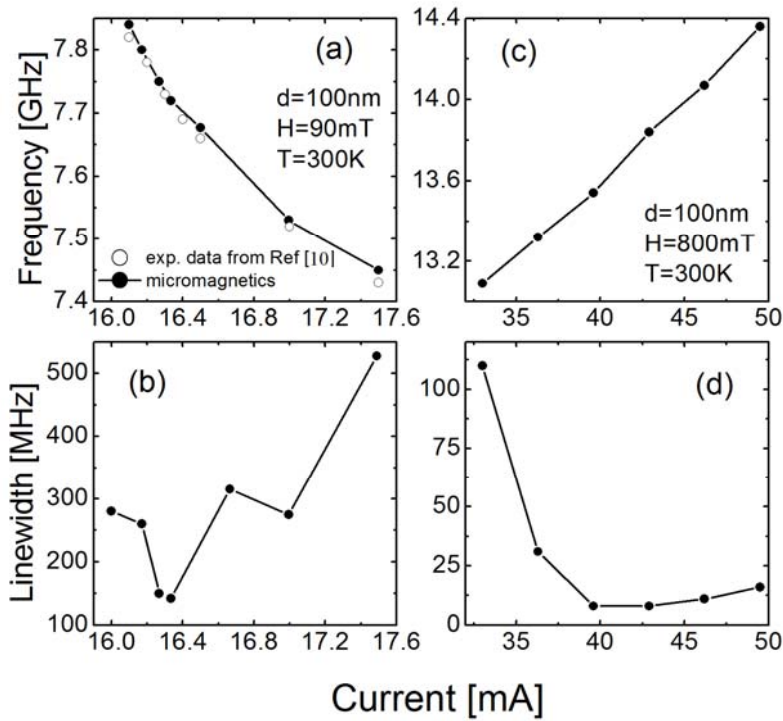

FIG. 3 (a) A comparison between experimental oscillation frequencies from Fig. 2(d) of Ref[10] (empty circles) and the micromagnetic computations (solid line with circles), and (b) predicted micromagnetic linewidths (in-plane $H$=90mT, $d$=100nm and $T$=300K) as a function of the current. (c) Predicted oscillation frequencies and (d) linewidths as a function of the current for an out of plane field $H$=800mT, $d$=100nm and $T$=300K.

In addition, we test the prediction of this model directly to the experimental data by performing the micromagnetic simulations at $T$=300K. The hysteretic behavior of the critical current disappears at room temperature, and the critical currents as a function of the field are coincident to the $I_{OFF}$ of Fig. 1(d) and in agreement to the trend measured in the experimental data (compare the $I_{OFF}$ curve in Fig. 1(a) to Fig. 4(a) in Ref. [10]). Fig. 3(a) shows a comparison between the experimental oscillation frequencies (red line in Fig. 2 (d) of Ref.[10]) and the ones computed micromagnetically as a function of the current ($H$=90mT, $d$=100nm). As can be observed, a good agreement is found pointing out that the approximations used to simplify Eq. 1 into Eq. 2 are consistent within this experimental framework. In this study, we considered the magnetization dynamics in high field regime ($H \geq$ 90mT), where a single mode is excited, and the devices can be used as harmonic microwave oscillators. At lower field, the oscillation is characterized by multimode power spectra similarly to what observed in[12,29]. In this region the Oersted field has a key role as also recently demonstrated in Spin-Transfer-Torque (STT) oscillators,[30,31] however those results are out of the aim of this work and will be discussed elsewhere. Fig. 3(b) displays the linewidth as a function of the current ($H$=90mT, $d$=100nm), a minimum of 142MHz is observed at $I$=16.3mA. The lineshape of the power spectrum is well approximated by a Lorentzian function and the linewidths have been computed as the full width at half maximum of the Lorentzian fitting of the power spectra.

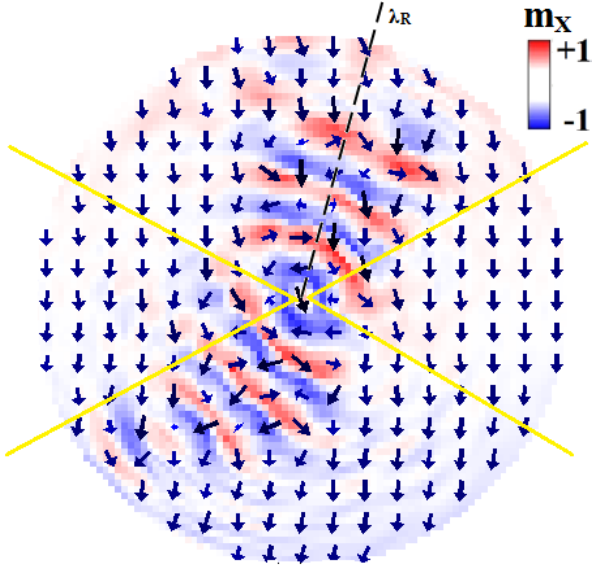

FIG. 4: Example of spatial distribution of the magnetization as computed by means of micromagnetic simulations for out-of-plane field $H$=800mT ($I$=37mA) (the color is related to the $x$ component of the magnetization, the arrows indicate the in-plane component of the magnetization). The dashed line shows the ideal path for the estimation of the wavelength $\lambda_R$. (multimedia view).

The same parameters have been used to study the behavior in out-of plane fields ($d$=100nm). In particular for fields larger than 700mT, our results predict a qualitative different behavior compared to the case of the in-plane configuration. Fig. 3(c) shows the oscillation frequencies as a function of the current for $H$=800mT ($d$=100nm), the tunability has a different sign than the case of in-plane fields, it changes from red to blue shift ($df/dI \approx$ 120MHz/mA). Our results indicate that a propagating mode is excited and a reduction of one order of magnitude of the linewidth is also predicted (compare Fig. 3(b) to Fig.3(d)). One reason for this can be attributed to the different origin of the excited mode. For a non-linear oscillator, the linewidth is inversely proportional to the magnetic volume involved in the dynamics,[32] and here the magnetic volume of the propagating mode is larger than the one of the localized mode. Differently from what observed in point contact geometries, where exchange-dominated cylindrical spin-wave modes are excited (namely linear-Slonczewski mode), here the profile of the propagation wave is strongly asymmetric.[33] This asymmetry is related to the spatial distribution of the current density. In other words, the current density near the contact region is large enough to compensate the Gilbert damping, exciting a self oscillation. On the other hand, the current density spreads (see Fig. 1(b)) when flowing between the two Gold contacts and from a critical distance it gives rise to a negative damping which compensates the Gilbert damping only partially. An additional source of asymmetry is the Oersted field and the *y*-component of the torque from spin-Hall effect which is an odd function if considering the center of the disk as the origin of the Cartesian coordinate system $\sigma_y(x,y) = \sigma_y(x,-y)$. Fig. 4 shows an example of the snapshot of the magnetization, where a clear asymmetric path of propagation can be observed. We estimated a wavelength $\lambda_R = 325 \pm 5$nm along the dashed line as displayed in Fig. 4.

The geometry of this spin-Hall oscillator can be qualitatively compared to a STT nanocontact oscillator (STNO). In an STNO, a localized spin-polarized current density is injected via a nano-aperture in an extended ferromagnet, the excited mode depends on the direction of the external field, and in fact a localized "Bullet" and a linear propagating Slonczweski mode are excited for in-plane and out-of-plane configuration respectively.[13, 29, 34] The tunability of the oscillation frequency also changes as a function of current from red to blue shift. However some differences can be underlined. In STNOs, the bullet mode is characterized by a uniform precession (same phase) of the spins below the nanocontact, differently the localized mode observed here presents spins which oscillate at the same frequency but with different phase (for instance see Fig. 2(d) spin #3). This difference is related to the non-uniform torque due to the spatial configuration of the current density, in fact this dephasing is more evident for oscillator with Gold contacts at larger distance. Also the spatial structure of the propagating mode is different. While the STNO can be seen as two-dimensional system, in which magnetic excitations can propagate in the whole plane,[33] here the propagation is asymmetric in the plane with the advantage that the spin waves can propagate for a longer distance compared to the STNO. While field tunable radiation patterns have been measured for STNO,[35] here the direction of the field plays a crucial role only for the nature of the excited spin wave being the polarization of the spin current independent on the field itself. In other words, to observe dynamical precession of the magnetization, the direction of the current, the in-plane component of the out-of plane field and the out of plane direction should form a right hand set of orthogonal vectors.

While the critical current are comparable for in-plane field, the spin-Hall oscillator has the disadvantage to need larger currents (30mA compared to 10mA) to excite propagating spin waves. There are at least two reasons for this. Firstly, the Pt has a spin-Hall angle of 0.08 which is smaller than typical values of spin-polarization $\approx 0.35$ for Py in STNO. Secondly, the negative damping due to the spin-Hall effect has the polarization always directed in the plane, while an out-of-plane field tilts the STNO polarizer out of-plane. This out of plane component can reduce the critical current significantly.

In summary, this letter introduces a micromagnetic framework able to describe recent experiments of magnetization dynamics driven by an in-plane current in heavy metal/ferromagnet bilayer. Similarly to that observed in STNO, it is possible to identify two different regimes of dynamical behavior, localized and propagating modes for in-plane and out-of-plane field direction, respectively. For in-plane fields, the oscillation frequency and the spatial distribution of the excited modes are in agreement with the experimental data as reported in[10]. For out-of plane fields, our findings show that this device geometry is a possible candidate for nanoscale spin-wave emitters for magnonic applications where spin waves are used for transmission and processing information on nanoscale.[13,36,37] Although the critical currents to excite propagating spin waves in spin-Hall oscillators are larger than the ones in STNOs, we believe that can be reduced by optimizing materials and geometrical properties, for instance, by considering the giant spin-Hall angle of Tungsten (W) in W/CoFeB bilayer which is at least three times larger than the one measured in Pt/Py.[38]

This work was supported by project MAT2011-28532-C03-01 from Spanish government and MIUR-PRIN 2010–11 Project 2010ECA8P3 'DyNanoMag'.

REFERENCES

[1] I. M. Miron, K. Garello, G. Gaudin, P-J. Zermatten, M. V. Costache, S. Auffret, S. Bandiera, B. Rodmacq, A. Schuhl, and P. Gambardella, Nature 476, 189-193, 2011.

[2] L. Liu, C-F. Pai, Y. Li, H. W. Tseng, D. C. Ralph, R. A. Buhrman, Science, 336, 555-558, 2012.

[3] P. P. J. Haazen, E. Murè, J. H. Franken, R. Lavrijsen, H. J. M. Swagten, B. Koopmans, Nature Materials, 12, 299-303, 2013.

[4] S. Emori, U. Bauer, S.-M. Ahn, E. Martinez, G. S. D. Beach, Nature Materials, 12, 611–616, 2013.

[5] K.-S. Ryu, L. Thomas, S.-H. Yang, S. Parkin, Nature Nanotechnology, 8, 527–533, 2013.

[6] D. Bhowmik, L. You, S. Salahuddin, Nature Nanotechnology, 2013 (published online doi:10.1038/nnano.2013.241).

[7] J. E. Hirsch, Phys. Rev. Lett. 83, 1834-1837, 1999.

[8] L. Liu, O. J. Lee, T. J. Gudmundsen, D. C. Ralph, and R. A. Buhrman, Phys. Rev. Lett. 109, 096602, 2012.

[9] C. O. Avci, K. Garello, I. M. Miron, G. Gaudin, S. Auffret, O. Boulle, and Pietro Gambardella, Appl. Phys. Lett. 100, 212404, 2012.

[10] V. E. Demidov, S. Urazhdin, H. Ulrichs, V. Tiberkevich, A. Slavin, D. Baither, G. Schmitz, S. O. Demokritov, Nat. Mater. 11, 1028-1031, 2012.

[11] L. Liu, C-F. Pai, D. C. Ralph, R. A. Buhrman, Phys. Rev. Lett. 109, 186602, 2012.

[12] R. H. Liu, W. L. Lim, and S. Urazhdin, Phys. Rev. Lett. 110, 147601, 2013.

[13] S. Urazhdin, V. E. Demidov, H. Ulrichs, T. Kendziorczyk, T. Kuhn, J. Leuthold, G. Wilde & S. O. Demokritov, Nature Nanotechnology 9, 509–513, 2014.

[14] M. Madami, S. Bonetti, G. Consolo, S. Tacchi, G. Carlotti, G. Gubbiotti, F. Mancoff, M. A. Yar, and J. Åkerman, Nature Nanotech. 6, 635, 2011.

[15] J. C. Slonczewski, J. Magn. Magn. Mater. 159, L1, 1996.

[16] T. Jungwirth, J. Wunderlich, K. Olejník, Nat. Mater. 11, 382-390, 2012.

[17] K. Garello, I. Mihai Miron, C. O. Avci, F. Freimuth, Y. Mokrousov, S. Blügel, S. Auffret, O. Boulle, G. Gaudin, P. Gambardella, Nature Nanotechnology 8, 587–593, 2013.

[18] E. Martinez and G. Finocchio, IEEE Trans. Magn. 49, 3105, 2013; E. Martinez, G. Finocchio, L. Torres, and L. Lopez-Diaz, AIP Adv. 3, 072109, 2013.

[19] G. Finocchio, M. Carpentieri, E. Martinez, B. Azzerboni, Appl. Phys. Lett. 102, 212410, 2013.

[20] S. Coco, A. Laudani, IEEE Transactions on Magnetics, 37, 3104-3107, 2001.

[21] X. Fan, J. Wu, Y. Chen, M. J. Jerry, H. Zhang, J. Q. Xiao, Nature Comm., 4, 1799, 2013.

[22] No differences in the dynamical response of the system in terms of oscillation frequency, linewidth and spatial distribution of the modes have been observed within this geometrical approximation.

[23] W. F. Brown, Jr. Phys. Rev. 130, 1677, (1963).

[24] D. M. Apalkov, P. B. Visscher, Phys. Rev. B 72, 180405(R), 2005.

[25] G. Finocchio, I. N. Krivorotov, X. Cheng, L. Torres, and B. Azzerboni, Phys. Rev. B 83, 134402, 2011.

[26] Z. Zeng, G. Finocchio, and H. Jiang, Nanoscale 5, 2219–2231, 2013.

[27] R. D. McMichael and M. D. Stiles, J. Appl. Phys. 97, 10J901, 1999.

[28] L. Torres, L. Lopez-Diaz, E. Martinez, G. Finocchio, M. Carpentieri, B. Azzerboni, J. Appl. Phys. 101, 053914, 2007.

[29] H. Ulrichs, V. E. Demidov, S. O. Demokritov, Appl. Phys. 104, 042407, 2014.

[30] R. K. Dumas, E. Iacocca, S. Bonetti, S. R. Sani, S. M. Mohseni, A. Eklund, J. Persson, O. Heinonen, J. Åkerman, Phys. Rev. Lett. 110, 257202, 2013.

[31] G. Consolo, G. Finocchio, G. Siracusano, S. Bonetti, A. Eklund, J. Åkerman, and B. Azzerboni, Jour. of Appl. Physics, 114, 153906, 2013.

[32] J.-V. Kim, V. Tiberkevich, A. N. Slavin, Phys. Rev. Lett. 100, 017207, 2008.

[33] J. C. Slonczewski, J. Magn. Magn. Mater. 195, L261-L268, 1999.

[34] S. Bonetti, V. Tiberkevich, G. Consolo, G. Finocchio, P. Muduli, F. Mancoff, A. Slavin and J. A. Akerman, Phys. Rev. Lett., 95, 237201, 2010.

[35] V. E. Demidov, S. Urazhdin, S. O. Demokritov, Nature Mater. 9, 984, 2010.

[36] H. Ulrichs, V. E. Demidov, S. O. Demokritov, S. Urazhdin, Appl. Phys. Lett. 100, 162406, 2012.

[37] S. Bonetti, J. Akerman, Magnonics Topics in Applied Physics, Springer Vol. 125, pp 177-187, 2013.

[38] C-F. Pai, L. Liu, Y. Li, H. W. Tseng, D. C. Ralph, and R. A. Buhrman, Appl. Phys. Lett. 101, 122404, 2012.